\documentclass{iau} 
\usepackage{graphicx}

\title[Resolved stellar populations with ELT-HARMONI] 
{From the inner Milky Way to Local Volume galaxies: resolved stellar populations with ELT-HARMONI}

\author[Oscar A. Gonzalez \& Giuseppina Battaglia]   
{Oscar A. Gonzalez$^1$
 \and Giuseppina Battaglia$^2$}

\affiliation{$^1$UK Astronomy Technology Centre, \\ Blackford Hill,
Edinburgh, EH9 3HJ, UK \\ email: {\tt oscar.gonzalez@stfc.ac.uk} \\[\affilskip]
$^2$Instituto de Astrofisica de Canarias \\ Calle via Lactea s/n, 38205 La Laguna, Tenerife, Spain
\\email: {\tt  gbattaglia@iac.es}}

\pubyear{2018}
\volume{347}  
\jname{EArly Science with ELTs (EASE)}
\editors{N. Przybilla, K. Pollard \& A. Calamida, eds.}
\begin{document}

\maketitle

\begin{abstract}
We discuss the predicted performance of the HARMONI spectrograph and the ELT in the context of two specific science cases: resolved stellar populations of Local Volume galaxies and Galactic archaeology in dense environments. We have produced and analysed a set of simulated data-cubes using the HSIM software, mimicking observations across the giant elliptical galaxy Centaurus A and in the nuclear bulge of the Milky Way. We use our results to demonstrate the instrument’s capabilities to perform stellar absorption line spectroscopy in a large number of stars which will allow us to study the detailed kinematics and stellar population characteristics of these high density regions.
\keywords{Keyword1, keyword2, keyword3, etc.}
\end{abstract}

\firstsection 
\section{Introduction}

The spectroscopic study of resolved stellar populations to determine the chemo-dynamical properties of galaxies and their components is fundamental to unravel the complex formation and evolution of galaxies from a very detailed perspective, which goes well beyond the determination of the mean chemical and kinematic properties.  The study of evolved low-mass stars such as red giant branch (RGB) stars is of crucial importance: low-mass stars can have lifetimes longer than a Hubble time and act then as \textit{living fossils}; Population II stars formed any time from the earliest star formation epochs to 1-1.5 Gyr ago are all expected to evolve as bright RGB stars. 

With current instrumentation on 8m-10m class telescopes, (low/intermediate resolution) spectroscopy of resolved RGB stars is possible within our own Local Group. While this hosts a very sizeable sample of dwarf galaxies of different types, it does not host any large elliptical and only a handful of small (M33 and the Large Magellanic Cloud) and large disc galaxies (the Milky Way \& M31). Making such studies possible within a volume of 4Mpc is a fundamental step to understand the formation history of particular systems like a large elliptical galaxy (Centaurus A, hereafter CenA) and to contrast the evolution of Local Group disc galaxies to those in other environments (e.g. the Sculptor group). 

On the other hand, within the Milky Way (MW), the innermost regions have for very long remained inaccessible because of the large amounts of dust extinction. Near-IR spectroscopy has opened the possibility to partially access those regions. However, with more than 3 magnitudes of extinction even in K-band, the need for a 40m class telescope to reach low-mass RGB stars ($\rm K>17.5$) is unavoidable. Furthermore, the high stellar crowding of these regions requires high spatial resolution to resolve stars down to these faint magnitudes. 

\section{Simulations}

We produced stellar populations with star formation histories and chemical enrichment laws broadly mimicking those of the systems to be analysed with simulations.  Synthetic stellar populations were created using the IAC-STAR software (\cite[Aparicio \& Gallart 2004]{aparicio_gallart04}). Note that IAC-star does not allow the choice of alpha-enhanced isochrones, so the synthetic stellar population was produced with solar-scaled models.  Three random realizations of 100000 stars each were used. 

\subsection{CenA stellar populations}

For the “large early-type” representative of CenA,  the star formation activity was assumed to be short and concentrated at old times (the star formation rate Is assumed to be constant from 13 to 10 Gyr ago), with metallicities distributed in the range $-0.3 < [Fe/H] < +0.2$ $(0.01 < Z < 0.03)$; the halo of CenA is kwown to have a MDF peaking at $\rm [M/H] \sim -0.6$ (see \cite[Rejkuba et al. 2005]{rejkuba05}), but since we will be probing regions closer to the center we move towards larger metallicities. We create mock catalogues with the stellar population mix (number of stars of given magnitude and colour) expected for CenA at radius R/Reff = 1 and 2, where Reff is the effective radius,  over a field of view of $6.4" \times 4.5"$. 

In order to estimate the expected stellar population mix in that given area, we adopt a distance modulus $\rm m-M= 27.92$ and a de Vaucouler profile with V-band $\rm \Sigma_{eff} = 22.15 mag/arcsec^2$; then $\rm \Sigma(R)= 8.327 \times [(R/Reff)^{(1/4)} -1 ] + \Sigma_{eff}$. Finally note that for this case, we neglect the effect of extinction. This is more or less equivalent to stay clear of the dust lane seen along the projected major axis of CenA.

\subsection{Galactic centre stellar populations}

For the case of the Galactic centre regions, we simulated two episodes of star formation peaking at 4 and 10 Grys, with metallicities distributed in the range $-1.5 < [Fe/H] < +0.2$. Although this would not necessarily be representative of the actual stellar populations in the GC regions, what is most important is that the population synthesis is suitable enough to reproduce the location of young/intermediate/old populations in the colour-magnitude diagram, thus providing a good test-bed for the detectability and mapping of such stellar population using HARMONI and, in particular, our ability to trace old stellar populations in this region.

The density of simulated stars in the field of view is adapted to reproduce a stellar density of 10 $\rm stars/arcsec^2$ in the magnitude range $17.5 <Ks <18.5$ within the inner two arcsec from the Galactic centre (\cite[Gallego-Cano et al. 2017]{Gallego18}). To compute the observed magnitudes we adopt a mean distance of 8 kpc (m-M=14.5 mag) and a reddening E(J-K)=5.68 mag (\cite[Gonzalez et al. 2018]{gonzalez08}). 

\subsection{Synthetic stellar spectra and creation of input cube}

A synthetic spectrum in the H-band was created for each star of the input catalogue. This was done by constructing a grid of spectra in the Effective Temperature-surface gravity-metallicity space and associating each star to the spectrum with the closest stellar parameters within the grid. The input spectra were obtained from the \cite[Coehlo et al. (2005)]{coehlo05} library. In each case, the flux of the synthetic spectra was normalized to the corresponding flux for reference wavelength at H-band, assuming a Vega magnitude scale for the magnitudes from the input catalogue. 

The input cube was created with a spatial sampling of 4mas and a spectral resolving power of $\rm R\sim30,000$, using the spatial distribution of the synthetic stellar population and their assigned spectra. Note that the FOV of the cube input cubes were limited to $\rm 1"\times1"$ in order to facilitate the computation time of the simulations.

\begin{figure}[]
\begin{center}
 \includegraphics[width=5.4in]{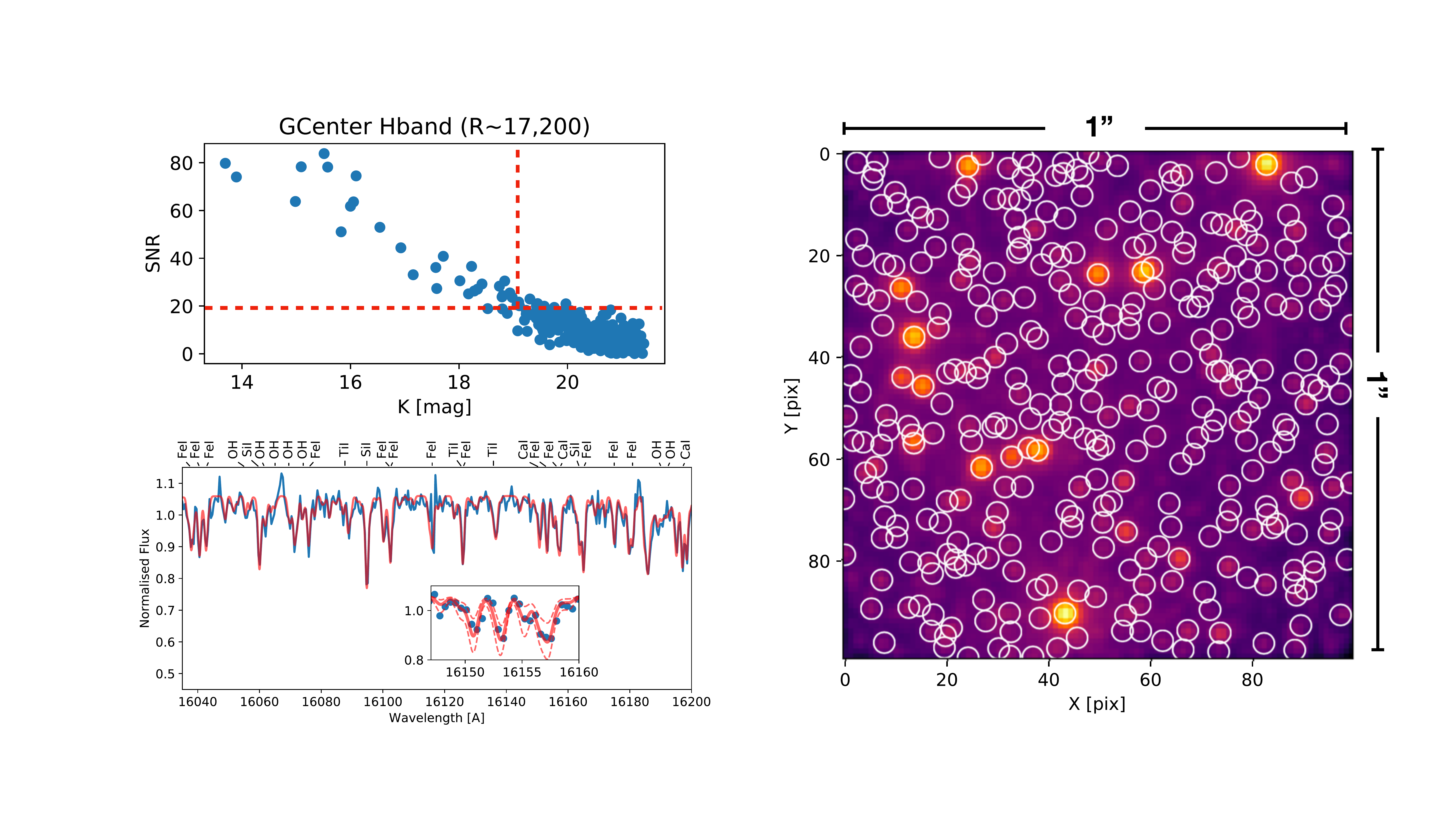} 
 \caption{White-light image of the simulated HARMONI cube with all sources extracted by PAMPELMUSE. Upper left panel shows the signal-to-noise ratio at central wavelength of the extracted spectra for all sources as a function of their K-band magnitude. The lower left panel shows an example of the extracted spectra at SNR=20 compared to the input spectrum (convolved to R=17200).}
   \label{fig1}
\end{center}
\end{figure}

\subsection{HSIM and extraction of simulated spectra}

The input cubes were fed to HSIM, the HARMONI simulation pipeline (\cite[Zieleniewski et al. 2015]{hsim15}), to simulate observations in average conditions with HARMONI. For CenA we selected the H-band grating providing intermediate resolution ($R\sim7,000$) spectra, and simulated a total exposure of 5h under $0.7"$ seeing and 1.2 airmass conditions. For the Galactic centre we used shorter exposure times of 1h using the H-band grating at high-resolution ($R\sim17,500$) under $0.7"$ seeing and 1.1 airmass conditions. All simulations were done using the 10 mas spatial sampling that provides a total field-of-view of $2.0"\times1.5"$ using single-conjugated adaptive optics (SCAO).

Source were extracted from the resulting cubes using PampelMuse (\cite[Kamman et al. 2013]{kamann13}). PampelMuse was developed to recover spectra from blended sources in MUSE data (e.g. \cite[Kamann et al. 2018]{kamann18} for an application to AO observations) but it can be nicely applied to extract sources from cubes of other IFU instruments. 

\section{Results}

\begin{figure}[]
\begin{center}
 \includegraphics[width=5.4in]{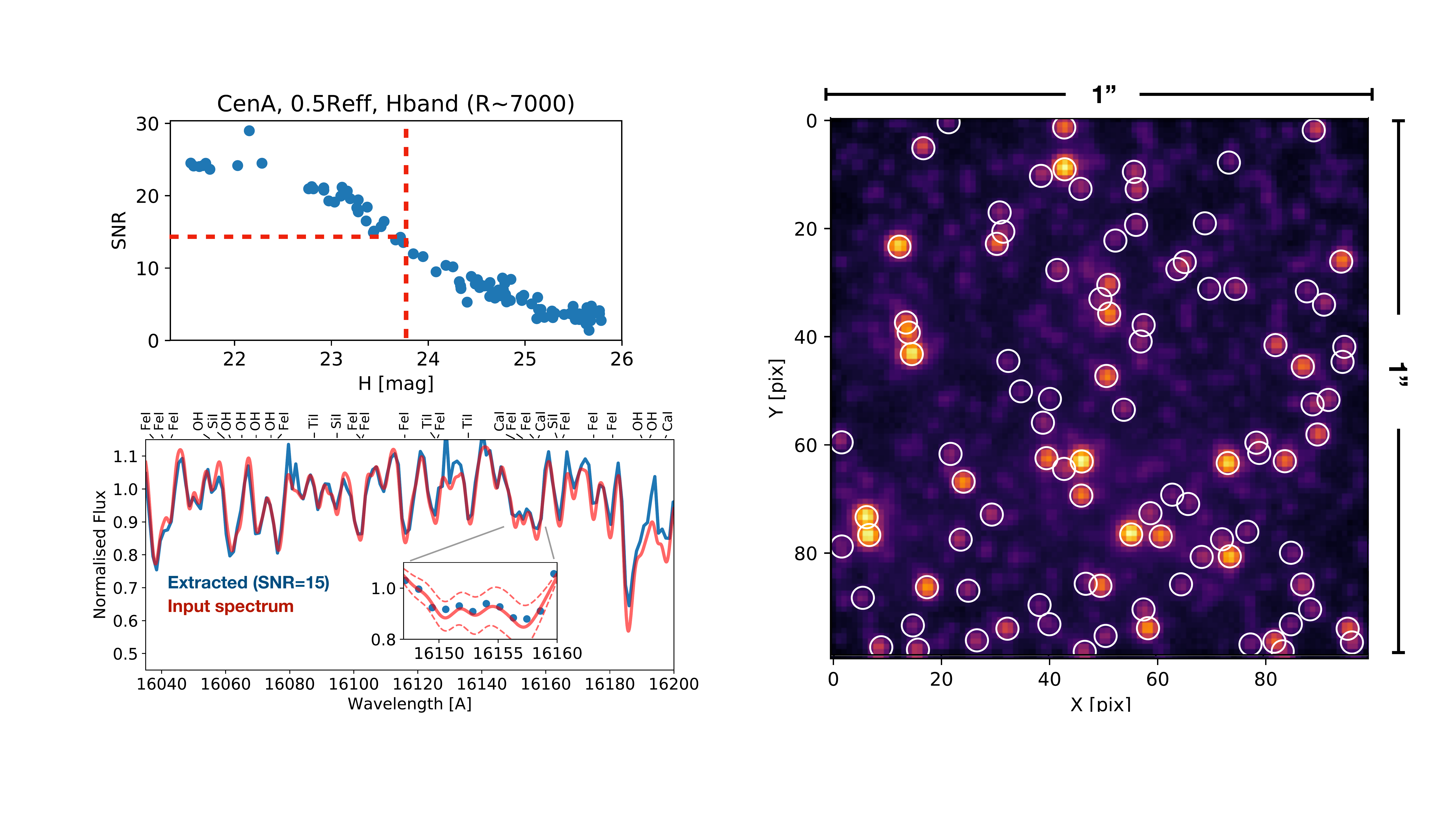} 
 \caption{Same as in Fig.1 for the case of CenA at 0.5$\rm R_{eff}$. The lower left panel shows an example of the extracted spectra at SNR=15 compared to the input spectrum (convolved to R=7000).}
   \label{fig2}
\end{center}
\end{figure}

The successfully extracted sources are shown in Fig.~\ref{fig1} and \ref{fig2} as well as their signal-to-noise ratio (SNR) as a function of magnitude. For the case of the Galactic centre we use K-band magnitudes to contrast the detection limits of HARMONI with the photometric studies of these regions. The K-band luminosity function of the Galactic centre region shows that the old population dominates at Ks=17.5 or fainter (\cite[Sch{\"o}del et al. 2017]{schoedel07}). By extending our simulations to the entire FOV of HARMONI at 10mas spaxels we show that it will be possible to obtain spectra for $\sim90$ stars at a $\rm SNR>20$ with a resolving power of R$\sim$17000 in 1hr exposure. In a one night (10hr) observation of the ELT, it would be possible to map a mosaic region of $5"\times10"$ and we predict a sample size of nearly 900 stars at $\rm SNR>20$. 

The predictions for the case of CenA at $ \rm R=0.5 R_{eff}$ are shown in Fig~\ref{fig2}. In this case, we show that in 10hr a sample of more than 100 stars can be extracted from the full HARMONI FOV (35 stars in the simulated $1"\times1"$) with a $\rm SNR>15$, which is sufficient for deriving individual radial velocities and metallicities, down to a magnitude of H=23.7.

Our results show that ELT-HARMONI can deliver a detailed characterization of the properties of large numbers of resolved low-mass stars in the vicinity of the galactic centre and in the dense regions of Local Volume galaxies out to 4 Mpc. 

\newpage
\begin{discussion}

\discuss{Bono}{Very interesting. How do you deal with Telluric Subtraction in Low Resolution mode?}
\discuss{Gonzalez}{At this stage we simply use the transmission cube which is one of the products from HSIM. This is a "perfect telluric". Our plan for the next stage of these simulations is to look into the telluric correction performances that we can expect from the real observations, including the known dependences to factors such as LSF variations, etc.}
\discuss{Whitelock}{Great talk. I think you might also find a few extreme AGB stars Rejkuba+ 2004 found Miras with H=19->21 in NGC5128. Perhaps there are odd ones in the Bulge too?}
\discuss{Gonzalez}{Thank you. Indeed there are Miras across these regions, some of which have been studied with other photometric surveys. However, in the case of the small HARMONI FOV it would be very rare to have those bright stars given their small densities, unless selected intentionally for such purpose.}

\end{discussion}

\end{document}